\documentclass[10pt]{article}
\usepackage{ifthen}
\newboolean{osa}
\setboolean{osa}{false}

\ifthenelse{\boolean{osa}}
{
\usepackage[OE]{express}
}
{
\usepackage{graphicx}
\usepackage{amsfonts}
\usepackage{amssymb}
\usepackage{hyperref}
}

\usepackage{amsmath}
\usepackage[utf8]{inputenc}
\newcommand*\conj[1]{\overline{#1}}

\begin{document}

\title{Using Convex Optimization of Autocorrelation with Constrained Support and Windowing for Improved Phase Retrieval Accuracy}

\ifthenelse{\boolean{osa}}{
\address{
\author{Alberto Pietrini,\authormark{1} Carl Nettelblad,\authormark{2,*}}
\authormark{1}Laboratory of Molecular Biophysics, Department of Cell and Molecular Biology, Uppsala University, Husargatan 3 (Box 596), SE-751 24 Uppsala, Sweden\\
\authormark{2}Division of Scientific Computing, Department of Information Technology, Science for Life Laboratory, Uppsala University, L{\"a}gerhyddsv{\"a}gen 2 (Box 337), SE-75105 Uppsala, Sweden
}
\email{\authormark{*}carl.nettelblad@it.uu.se} 
}
{
\author{Pietrini, Alberto \\ \texttt{alberto.pietrini@icm.uu.se} \and Nettelblad, Carl \\ \texttt{carl.nettelblad@it.uu.se}}
\maketitle
}



\begin{abstract}
In imaging modalities recording diffraction data, the original image can be reconstructed assuming known phases. When phases are unknown, oversampling and a constraint on the support region in the original object can be used to solve a non-convex optimization problem.

Such schemes are ill-suited to find the optimum solution for sparse data, since the recorded image does not correspond exactly to the original wave function. We construct a convex optimization problem using a relaxed support constraint and a maximum-likelihood treatment of the recorded data as a sample from the underlying wave function. We also stress the need to use relevant windowing techniques to account for the sampled pattern being finite.

On simulated data, we demonstrate the benefits of our approach in terms of visual quality and an improvement in the crystallographic R-factor from .4 to .1 for highly noisy data.
\end{abstract}

\ifthenelse{\boolean{osa}}{
}
{}

\ifthenelse{\boolean{osa}}{
\bibliography{coacs}{}
}{}
\bibliographystyle{osajnl}

\section{Introduction}

In imaging applications ranging from astronomy to molecular biology, diffraction-based imaging techniques have been demonstrated to be useful. In general, the diffraction from an optically thin object in the far field at low angles can be approximated as the Fourier transform of the scattering power of a projection of the original object. Thus, to reconstruct this projection, the inverse Fourier transform should be applied to the collected image. However, imaging detectors tend to only record light intensity (amplitude squared), losing the complex phase of the underlying wave.

In crystallographic applications, common in imaging of biological particles like proteins, this \emph{phase retrieval} problem becomes under-constrained, based on the physical data alone. The so-called Bragg peaks representing constructive interference between all unit cell repeats will not, in general, constrain the content of the unit cell to a single representation, i.e. a single phase assignment. By introducing knowledge of possible biological structures in different ways, the problem still becomes tractable \cite{crystalphasing}.

For isolated particles, phase retrieval based on oversampling relative to a support constraint has become the method of choice. The resulting equation system is non-linear, and solving it in a minimum residual sense results in a non-convex optimization problem. Thus, most solution methods do not make claim on providing global optima, but try to balance finding optima and exploring a reasonable portion of the domain. Popular solution methods use variations of alternating projection techniques, such as Hybrid-Input Output (HIO, \cite{hio}) and Relaxed Averaged Alternating Reflections (RAAR, \cite{raar}). A concise comparison of various iterative schemes and their similarities can be found in \cite{marchesini}.

Several of these iterative methods were developed for conditions of relatively high signal, with no or only moderate noise. This is very different from a sparse imaging regime with pixel detectors able to record individual photons, which is now the reality in e.g. FXI (Flash X-Ray Imaging), where individual biological particles are imaged in femtosecond-duration X-ray pulses delivered using X-ray free electron lasers, such as the Linac Coherent Light Source (LCLS \cite{lcls}). In addition, some parts of the image might be missing due to detector geometry, detector malfunction, or the geometry of the experiment itself. Despite these limitations, successful reconstructions have been reported for organic and inorganic samples alike \cite{nanorice,flashman,mimi,carboxy}.

In this communication, we introduce a relaxed version of the support constraint, based on the known relationship between the Fourier transform of the original wave, and the Fourier transform  of the observed quantity, the intensities. We show the convex character of the optimization problem of estimating maximum-likelihood intensities given a recorded sparse sampling of the diffraction pattern, a support, and a noise model. Convex optimization problems lack separate local optima and allow a wealth of existing methodology to be used to identify points arbitrarily close to the true global optimum. We name this method Convex Optimization of Autocorrelation with Constrained Support (COACS).

COACS has similarities to methods such as PhaseLift \cite{phaselift1,phaseliftsiam2015} and PhaseCut \cite{phasecut}. However, we deliberately disentangle finding optimum intensities from finding the phases. This, together with great care in the numerical implementation, seems to partially overcome the numerical stability concerns that led the authors in \cite{phaseliftsiam2015} to conclude that not only would PhaseLift be unsuitable for phase retrieval of oversampled particles with known support, but that the problem itself is intractable to numerical solution without adding further constraints.

We demonstrate a proof of concept solution method using a pre-existing software package for compressed sensing, TFOCS (Templates for First-Order Conic Solvers) \cite{tfocs}. We evaluate phase retrieval with and without COACS pre-processing based on TFOCS, realizing marked improvements.

In the remaining sections, we first describe our optimization problem in greater detail, followed by a detailed account of implementation concerns, experiments, results, and their evaluation.

\section{Intensity healing using COACS}
We will focus on the case of 2D imaging, since it is straightforward, yet rich enough to illustrate most of our points. The overall methodology generalizes to 3D, like the iterative phase retrieval used as the final step in \cite{3dmimi}, and for that matter to further dimensions.

Assume that the 2D projection being imaged is $\mathit{P}$. Assuming ideal conditions (a plane incident wave, a thin object, a far-field regime, and only considering low scattering angles), the scattered wave $\mathit{X}$ meeting a square detector satisfies the following relation:
\begin{equation}
\mathit{X} \propto \mathcal{F} (\mathit{P})
\label{fouriercont}
\end{equation}

Since actual photon detectors use discrete pixels covering a limited region in space, we will treat a discretized, finite case, where $\mathbf{P}$, $\mathbf{X}$ (matrices in bold) are 2D center-cropped discretizations of their continuous counterparts. For practical purposes, we can then consider the following version of eqn. \ref{fouriercont}:
\begin{equation}
\mathbf{X} \propto \mathcal{F} (\mathbf{P})
\label{fourierdisc}
\end{equation}

\begin{figure}[htbp]
\centering\includegraphics[width=12cm]{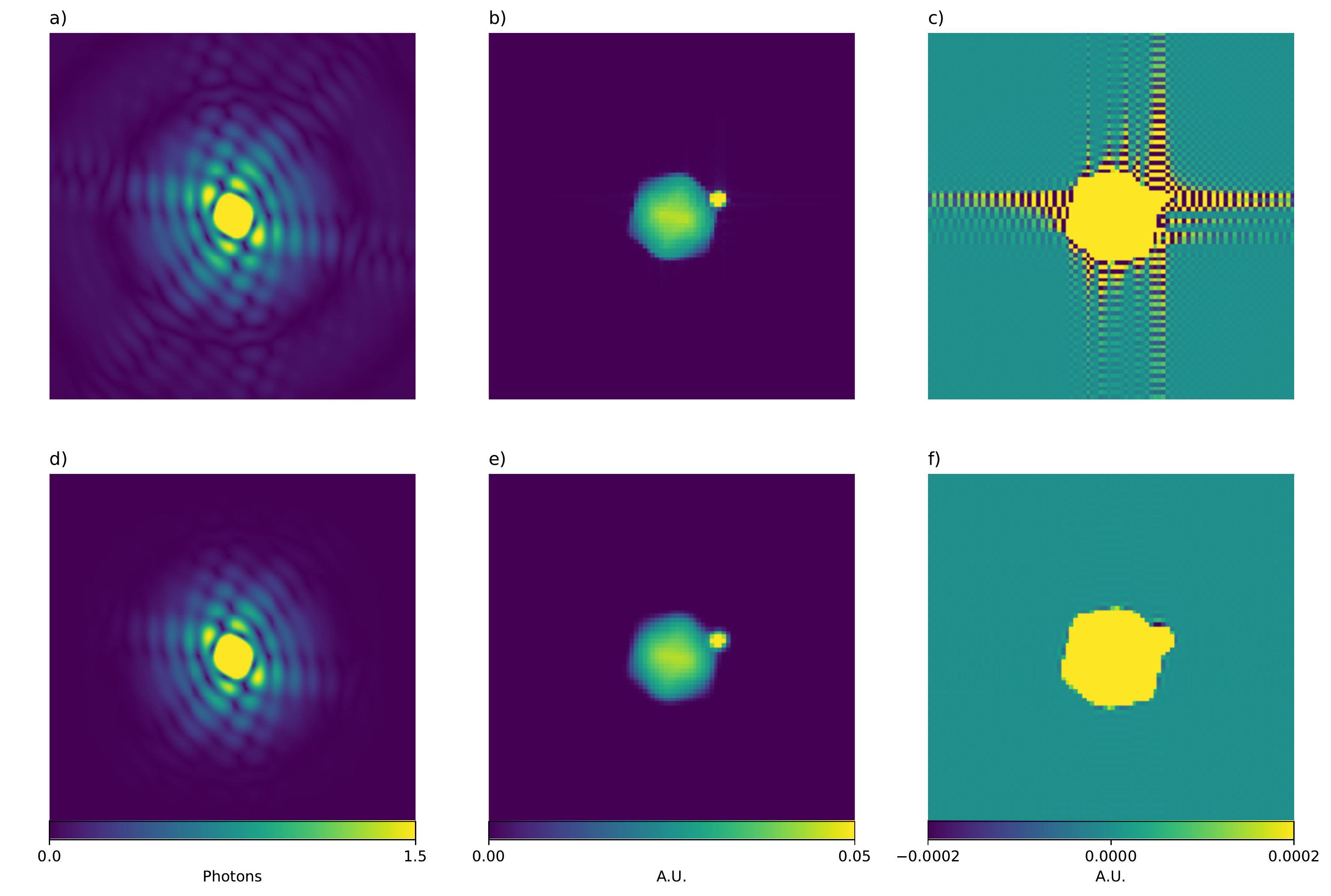}
\caption{Simulated particle with and without Hann windowing. a) Original simulated diffraction pattern with high-density sphere. b) Center crop of discrete Fourier transform of a), showing the particle. c) The image in b) with limited range to showcase artifacts outside of the object outline. d) The simulated pattern with Hann windowing applied, resulting in lower intensity away from the center. e) Discrete Fourier transform of pattern after windowing. Slight blur visible. f) The image in e) with limited range. Artifacts found in c) mostly absent.}
\label{cropped}
\end{figure}

However, these two formulations are not identical. The discrete Fourier transform assumes that the underlying object is periodic. Figure \ref{cropped} illustrates the effects of discretizing the infinite ideal diffraction pattern of our simulated test particle, with small but non-negligible artifacts permeating to the edge of the image. The physical processes that we model are based on eq. \ref{fouriercont}. To handle those properly in the discretized space, we extend eq. \ref{fourierdisc} with the Hann window, known from spectral techniques in signal processing. In 2D, the Hann window in real space amounts to a convolution with a 3x3 blurring stencil, and in the Fourier space a translated sinusoid curve dropping off to 0 at the edges. This ensures a representation where the high-frequency content smoothly goes to zero. Having no explicit window is equivalent to a sharp rectangular window, giving rise to the spectral leakage effects seen in Figure \ref{cropped}.

If we can assume a photon-counting detector with uniform quantum efficiency $r$, we will then have a sampled diffraction pattern in terms of an integer matrix $\mathbf{B}$, where:
\begin{equation}
\mathbf{B}_{i,j} = \mathit{Po}(r \mathbf{X}_{i,j} \conj{\mathbf{X}}_{i,j}) = \mathit{Po}(r|\mathbf{X}_{i,j}|^2)
\label{sampling}
\end{equation}

where $\mathit{Po}$ is a Poisson distribution with the rate parameter $\lambda$ of appropriate dimension identifying the mean (and variance). For high rates, $\mathbf{B}$ can be treated as a Gaussian, or even as an exact observation $\mathbf{B}_{i,j} = r |\mathbf{X}_{i,j}|^2$. Even a strong diffraction pattern will tend to have a characteristic structure of minima, similar to those from the most trivial slit experiment. Thus, even though the pattern as a whole is strong, the Gaussian or identity approximation might not hold for all pixels within the pattern.

\subsection{Support constraints}
Since a single particle is imaged in isolation, we can introduce a support constraint, where $\mathit{P}_{\mathit{S}^\complement}$, or $\mathbf{P}_{\mathbf{S}^\complement}$, is 0 (where $\complement$ signifies the complement). For phase retrieval, one can thus express this as:
\begin{equation}
\operatorname*{arg\,max}_{\{\mathbf{X} \mid \mathbf{P}_{\mathbf{S}^\complement} = 0  \} } P(\mathbf{X} \vert \mathbf{B})
\label{support1}
\end{equation}

Alternating-projection methods are very commonly used for phase retrieval. They were pioneered by Fienup \cite{hio}, and are still popular in practical applications use, as e.g. implemented in \cite{hawk}. The name comes from the fact that they alternatingly implement some variation on the support constraint, and the intensity (``Fourier'') constraint. The standard implementation is then to assume that the intensities are exact, rather than stochastic as given above. Both of these constraints can be varied, to better explore the full solution space and accelerate convergence \cite{hio,raar}, or in order to account for measurement error in the intensities \cite{thibaulterror}. More recent attempts include using the Alternating Direction Method of Multipliers for easily enforcing more general constraints in both spaces \cite{slacpaper}.

These attempts share the property that they do not claim global convergence. Recent advances apply semi-definite and convex programming approaches to the full phasing problem, finding low-rank factorizations for matrices in order to produce the phases, such as PhaseLift \cite{phaselift1} and PhaseCut \cite{phasecut}. These approaches claim to produce unique solutions under favorable conditions with a vanishing probability of not doing so, such as observing an object using alternating binary masks, or from unique angles randomly distributed over a sphere \cite{phaselift1}. While such conditions are possible to achieve in some situations, they would be challenging to transfer to the single particle X-ray diffraction experiments.

PhaseLift considers the rank-one matrix $p\conj{p}$, where $p$ would be the flattened vector from our 2D matrix $\mathbf{P}$. A megapixel image would thus give rise to a trillion-element matrix. It is possible to formulate an alternate operator based on the Fourier transform that goes from such representations to the diffraction pattern. The problem can then be solved in terms of satisfying the observed data with a suitable matrix in this space, while also minimizing the rank. A convex representation of that problem can be expressed as minimizing the trace.

The authors of PhaseLift note that while oversampling in theory gives a unique solution to the phasing problem, uniqueness is not equivalent to practical numerical attainability. In fact, they even claim, based on simulations including their own method as well as alternating projections \cite{phaseliftsiam2015} (original italics):
\begin{quotation}
The ill-posedness of the problem is evident from the disconnect between small
residual error and large reconstruction error; that is to say, we fit the data very well and yet
observe a large reconstruction error. \emph{Thus, in stark contrast to what is widely believed, our
simulations indicate that oversampling by itself is not a viable strategy for phase retrieval even
for nonnegative, real-valued images.}
\end{quotation}

This argument is supported by additional claims about numerical stability. We will argue that a new analysis, of the support constraint specifically, will allow a much more favorable numerical treatment.

\subsection{A support constraint for the Patterson function}
The quantity that is sampled to form the directly observable $\mathbf{B}$ is $\mathbf{Y} = \mathbf{X} \odot \conj{\mathbf{X}}$ (the elementwise Hadamard product with the conjugate, forming the absolute value squared). In crystallographic applications, the Fourier transform of $Y$ is referred to as the Patterson function. Using the convolution theorem, we can characterize this as:
\begin{equation}
\mathcal{F}(\mathbf{Y}) = \\
\mathcal{F}(\mathbf{X} \odot \conj{\mathbf{X}}) = \\
\mathbf{P} \ast \conj{\mathbf{P}}
\end{equation}

Here, $\ast$ indicates convolution. The existing support constraint \eqref{support1} for $\mathbf{X}$ can then be formulated as a (weaker) support constraint for $\mathbf{Y}$:
\begin{equation}
\operatorname*{arg\,max}_{\mathbf{Y}, \mathcal{F}(\mathbf{P} \ast \conj{\mathbf{P}})_{(\mathbf{S} \ast \mathbf{S})^\complement} = 0} P(\mathbf{Y} \vert \mathbf{B})
\end{equation}
\begin{equation}
\operatorname*{arg\,max}_{\mathbf{Y}, \mathcal{F}(\mathbf{\hat P})_{ \mathbf{\hat S}^\complement} = 0} P(\mathbf{Y} \vert \mathbf{B})
\label{support2}
\end{equation}

Based on the structure of the probability distribution function in \eqref{sampling}, \eqref{support2} is a likelihood-optimization problem with a linear transform into independent variables. The previous quadratic term for $\mathbf{X}$ has vanished, at the possible expense of the self-convoluted version of $\mathbf{S}$, $\mathbf{\hat S}$, imposing a less strict constraint on the solution. The self-convolution effect of treating the intensity directly has also led some authors to refer to this as the autocorrelation function.

\section{Implementation}
If the noise structure would have been Gaussian, the resulting problem from \eqref{support2} would have amounted to an ordinary least squares problem. A solution could be determined by solving the over-determined homogeneous linear equation system based on the Fourier operator for those pixels that are in the complement of the self-convoluted support. For a practical solution, one will however also need to add the constraint of all intensities being non-negative, which necessitates specialized convex optimization solvers, rather than the most basic least squares routines. Negative intensities would be impossible to transform back to amplitudes in the original phasing problem.

For a long time, so-called interior point methods have been a method of choice for many families of convex optimization problems \cite{interiorreview}, with free \cite{solverfree} as well as commercial \cite{solvercomm} solvers, and more high-level modeling environments \cite{cvx}. These methods explicitly form the Hessian of a problem being solved, with a number of elements equivalent to the square of the number of variables. They tend to rely on sparseness in the dependencies between variables. Unfortunately, whereas the support constraint is one type of sparsity, the Fourier operator itself is dense. In some solver implementations, the operator would also have to be formed explicitly as a matrix, rather than the far more efficient Fast Fourier Transforms. The benefit of interior point methods is that they tend to converge very quickly, in terms of the number of iterations.

We have instead chosen to rely on first-order methods, where only the gradient is computed. The TFOCS \cite{tfocs} package is a Matlab toolbox, and is presented by the author as a set of templates from which one can build tailored solvers. There are a couple of different canonical forms for formulating problems in TFOCS, and we have used the one that corresponds to the main function call \texttt{tfocs}:
\begin{eqnarray}
\phi(\mathbf{X}) = f(\mathcal{A}(\mathbf{X}) + b) + h(\mathbf{X}) \\
\mathbf{X}_{\mathrm{opt.}} = \operatorname*{arg\,min}_{\mathbf{X} } \phi(\mathbf{X})
\end{eqnarray}

The linear operator $\mathcal A$ does not need to be implemented as a matrix, but can be expressed in code, assuming it satisfies certain specific criteria. We can thus rely on an actual efficient Fourier transform implementation. The function $f$ needs to be smooth and differentiable, with an explicit implementation of the gradient. $h$ on the other hand, only has to be prox-capable, meaning one can (quickly) solve:
\begin{equation}
\Phi_h(\mathbf{X}, t) = \operatorname*{arg\,min}_{\mathbf{Z} } h(z) + \frac{1}{2} t^{-1} ||\mathbf{Z}-\mathbf{X}||^2_2
\end{equation}

In our context, $f$ corresponds to the noise model, while $h$ represents the support constraint. Both of these are implemented as custom functions, for numerical accuracy reasons outlined in the next section.

\section{Numerical concerns}
As noted by the authors in \cite{phaseliftsiam2015}, numerical stability and accuracy concerns are of great importance in the phase retrieval application, especially when relying solely on the  oversampling of a single pattern. Although the point of doing a separate step of intensity healing is to make the phase-retrieval problem more well-posed, these numerical concerns are of importance for intensity healing as well. The intensity in a diffraction pattern of, say, a uniform sphere, will decay in a manner roughly proportional to $q^4$, where $q$ is the scattering angle \cite{sphere}. The dynamic range needed for a problem in intensity space is higher than in the amplitude space, due to the quadratic relationship between the two. Since the Fourier transform adds terms for all frequencies for each resulting pixel, and terms are of varying sign, loss of precision due to terms of opposite sign canceling out can be paramount. In any iterative approach, the numerical accuracy needs not only hold for the final result, but for properly evaluating the change undertaken between iterates.

The steps outlined below try to consistently reduce the chance of:
\begin{itemize}
	\item Truncation error due to very small terms being added to large terms.
	\item Loss of precision due to large negative values being added to positive values of almost equal magnitude, or vice versa.
    \item Very small steps due to the optimal solution being located at the border of the allowed domain.
\end{itemize}

The specific places where this has been taken into account can be summarized as:
\begin{itemize}
	\item \textbf{Lipschitz backtracking structure} in TFOCS
	\item \textbf{Translation of the solution variables and customized probability function}
    \item \textbf{Relaxed quadratic barrier and accelerated continuation}
    \item \textbf{Window-aware support constraints}
\end{itemize}

It should be noted that without these steps, one arrives at a seemingly mathematically equivalent problem formulation that still results in erratic numerical behavior in practice. The specific nature of these steps also relate to design choices in the existing TFOCS solver, including the fact that differences between successive iterates are computed, and that the objective function is handled as a scalar, rather than considering the individual per-pixel contributions. On the other hand, these properties are in no way unique to TFOCS.

\subsection{Lipschitz backtracking structure}
TFOCS can be described as an alternating-projections method tracing modified gradients, giving far accelerated convergence compared to a fixed-step gradient descent. For more details on the overall design of the package, we refer to \cite{tfocs,tfocsug}. One crucial aspect of the algorithm is a backtracking approach to identify the proper step size. The backtracking is common to the various specific solver schemes implemented in TFOCS. The backtracking is using a local, adaptive estimate of a Lipschitz constant for the gradient of the objective function.

As elaborated upon in \cite{tfocs}, there are multiple ways to estimate a bound on the Lispchitz constant based on locally evaluated gradients and function values. Since the problem is inherently related to evaluating the difference in function value between two points, there is a great risk for loss of precision. The original TFOCS approach is therefore to use two bounds, one of which will give less conservative estimates, while also being more sensitive to the details of floating-point math, and another that is safer.

The TFOCS code automatically detects whether the relative difference in function value between the two points is small, and then switches to the more conservative bound. The conservative method will then be used until the next ``restart'' of the TFOCS algorithm.

We have made two modifications to this logic. The first modification is to move the evaluation point for choosing which bound formulation to use. The original TFOCS code performs one step and then, retrospectively, checks whether the difference was small. With this design, a single step in the iteration process might be taken using an accuracy-deficient Lipschitz estimate. This turned out to be enough to cause drastic divergence for us in some scenarios. We instead put this test earlier. We also allow the iteration process to leave the conservative regime, without a restart. Furthermore, we consider not only the difference in function values $|f(x) - f(y)|$, but also the difference $|x - y|$. If the step itself is small relative to the magnitude of the values, truncation errors as well as loss of precision are possible.

The total change is less than $20$ lines of code in a single function in the package. It has been submitted to the authors as a suggested change. This, together with using the ``no-regress'' restart feature gives adequate optimization performance without divergence or singularity issues.

\subsection{Translation of the solution variables and customized probability function}
The Fourier transform is a linear operator. Thus, the relationship between $\mathbf{Y}$ and $\mathbf{\hat P}$ can be decomposed:
\begin{eqnarray}
\mathbf{Y} = & & \mathcal{F}( \mathbf{\hat P}) \\
\mathbf{Y} = & \mathbf{Y^0} + \mathbf{Y^\star} = & \mathcal{F}( \mathbf{\hat P^0}) + \mathcal{F}(\mathbf{\hat P^\star})
\end{eqnarray}

In this case, the optimization variable for the first-order method in TFOCS can be $\mathbf{Y^\star}$. With $\mathbf{Y^0}$ chosen properly, such as the resulting solution from a previous run of the TFOCS algorithm (the previous ``outer iteration''), the norm of $\mathbf{Y^\star}$ should be much smaller. This change reduces the effects of truncation errors in the gradient steps in individual iterations, including the probability of resorting to the conservative Lipschitz bound discussed above. It also reduces the effects of loss of precision due to large values of opposite sign canceling out within the Fourier transform. The effects do still exist on a global level, but they do not hamper the evaluation of objective values, constraint satisfaction, and search direction \emph{locally} during the iteration process. The equivalent translation is not possible in the phasing problem, since any changes in the phase estimates affect the full amplitudes, whereas an additive decomposition is possible in the intensity space.

To fully realize the benefits of translation, the corresponding shift needs to be done in the computation of the $f$ term of the objective function as well. We have implemented a custom log Poisson implementation, which computes the change in objective function value between $\mathbf{Y^0}$ and $\mathbf{Y}$ using $\mathbf{Y^\star}$ as input. To further accelerate convergence, especially for pixels with zero observed photons (where the log-Poisson optimum lies exactly at 0, the border of the non-negativity constraint), we replace the log-Poisson with a linear extrapolation below a border value $l$, and add a quadratic function scaled to still keep the minimum at 0 for the zero-photon case. Hence, negative values are allowed, but strongly penalized.

In our implementation, we have chosen to implement the windowing factor inside $f$, rather than in $\mathcal{A}$. This makes the effects of the windowing on the resulting gradients more visible (the gradients are scaled by the window squared). Since the reason to introduce $l$ is to ensure a smooth behavior amenable to optimization, it makes sense that the effective gradient should be of equal magnitude everywhere. Therefore, the actual value used for the barrier width is inverted by the window, resulting in a more lenient barrier in the high-frequency areas, but also requiring an overall lower value of $l$ to guarantee high accuracy at all angles. To stabilize the inverse process, the Hann window in amplitude space (which is squared in intensity space) has an extra term of $10^{-3}$ in order to avoid singularities.

\subsection{Relaxed quadratic barrier and accelerated continuation}
The extra quadratic barrier $l$ introduced in the previous section determines the sharpness of the non-negativity constraint. In our implementation, we initiate $l$ to a high value ($l = 4$). This value is then decreased by a factor of $.5$ repeatedly until reaching $2^{-46} \approx 10^{-14}$. Due to the Hann window scaling of $l$, this means an effective barrier of $10^{-8}$ at the edge of the image.

This scheme ensures a very quick convergence to the relaxed problem, followed by successive sharpenings in a continuation scheme, starting off from the previous end result. Let $\mathbf{X}_i$ denote the result from iteration $i$ (starting at 0, with $l = 2^{2-i}$). Since the major trend influencing the results should be the reduced fluctuations around 0, the initial guess for iteration $i$, when $i > 2$, will be $\mathbf{X}_{i-1} + 0.5 (\mathbf{X}_{i-1} - \mathbf{X}_{i-2})$, i.e. reproducing half the change that resulted from the previous halving of $l$. This acceleration scheme is similar to the ones more thoroughly investigated for the SCD model in \cite{tfocs}.

Naturally, the actual $\mathbf{X}$ values used are translated for the accuracy reasons already discussed. Each continuation step with a new $l$ value induces a rapid shift in the optimum, even after the acceleration adaption. Therefore, a second set of inner continuations is applied after a fixed number of individual inner iterations of the optimization scheme. In these inner continuations, a separate acceleration scheme is used after the second iteration, where the new starting point is displaced with an accumulation factor of $.9$. If at any point the resulting $\mathbf{X}$ estimate is found to have a shorter distance to the old end point than the new starting point, the acceleration is deemed unsuccessful and that inner iteration is restarted without acceleration. We have found this ad-hoc scheme to give very reasonable results with an acceptable time usage. The inner continuation loop is terminated when the change in the objective function, scaled by the value of $l$, falls below a pre-defined threshold.

\subsection{Window-aware support constraint}
The support constraint can be simply implemented as a projection to $0$ outside of the support. However, this might unnecessarily constrain the model in terms of inducing very small steps. We therefore ventured into employing a strong, but non-infinite, quadratic penalty, chosen to be $\frac{5e7}{l}$.

The quadratic penalty, as well as the original zero-projection constraint, has a drawback in terms of the spectral behavior. The sharp cutting at the edge of the support will induce high-frequency signals in the diffraction space, while we are already employing the Hann window to the original diffraction signal in order to reduce that signal. Therefore, we are using the same kind of two-level Hann window intensity space for the support penalty matrix in the Fourier space, which amounts to a gradual decrease in support penalty over a width of 5 pixels.

In practice, similar results were achievable without this addition, but the time usage was much higher. This can be understood by the fact that in our simulations, the support was in fact wider than the object. Therefore, the ideal solution is already zero in the border region. For a fully converged, ideal, solution, the two constraint formulations are therefore identical. The windowed version will only accelerate convergence by avoiding spectral leakage.

\section{Data and experiments}
Simulated diffraction data were pre-treated with our proof-of-concept implementation of COACS based on TFOCS, called \texttt{jackdaw} (available at \url{https://github.com/cnettel/jackdaw}).

Simulations were done using the Condor \cite{condor} package. The test particle chosen in the simulations consisted of an icosahedron, with a single very high-density sphere placed on a point off-center on one of the vertices. The shortest diameter between vertices was 20 pixels, imaged on a virtual 256x256 detector. The high-density sphere had a diameter of 4 pixels, with a much higher scattering power per volume than the icosahedron. The presence of a concentrated, markedly non-symmetric feature makes it possible to judge the presence of artifacts in resulting reconstructions, as well as the ability to resolve the feature at all. At the same time, our simulated sample is as a first approximation roughly spherical, which is appropriate for the types of biological particles one would want to image.

A total of 50 Poisson samplings were created of this particle. A central beamstop was simulated by making a 25x25 square stencil in the center missing. This  covered the majority of the central speckle for the icosahedron, resulting in on average 10,000 photons in the non-masked pattern. Intensity-healed patterns were computed using our COACS implementation for this pattern, using a trivial auto-correlation support consisting of a square with a side of 61 pixels. While a Shrinkwrap approach akin to what is common in phase retrieval \cite{shrinkwrap} would be possible, this choice was made to demonstrate the lack of specific assumptions on the structure of the sample, and how it would be possible to process a large number of shots of varying particles with little to no manual tuning.

Phasing was performed using the Python interface to \texttt{libspimage} with the GPU-based HIO phasing method, with the relaxation parameter $\beta = 0.9$, identically for COACSed (healed) and un-processed images. The support used was a square with a side of 31 pixels, i.e. slightly larger than the particle itself. 50,000 iterations of HIO were performed, followed by 10,000 iterations of pure ER to find a suitable local minimum. For each pattern, phasing was repeated 100 times, in order to control for the variations due to random initial data. The eventual reconstruction chosen for each pattern was the average of the 10 best out of these phasing results, chosen on the basis of the real space error, i.e. the norm of the remaining signal outside of the support.

\section{Results}
\begin{figure}[htbp]
\centering\includegraphics[width=12cm]{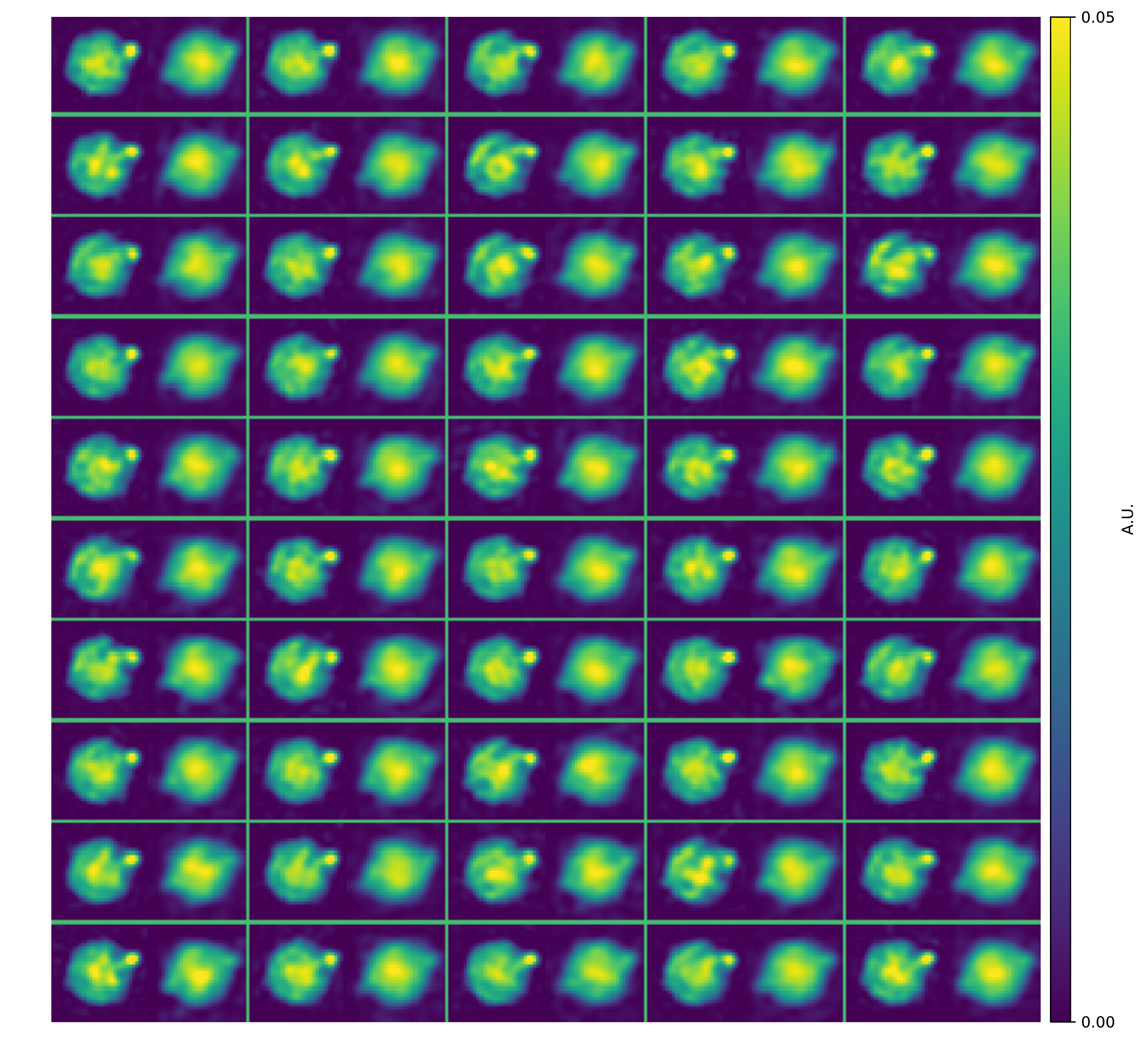}
\caption{50 phased reconstructions of sparse patterns based on the simulated particle. Pairs of phased healed pattern using COACS  (left) and phased original patterns (right). Average of 10 best individual phasings out of 100 replicates for each of these.}
\label{healingexamples}
\end{figure}

A qualitative comparison of the results possible using intensity healing are given in Figure \ref{healingexamples}, for the 50 patterns simulated. For each pair, the COACsed pattern is showed to the right. The high-density spherical feature on the edge is clearly visible in all healed reconstructions, and the contours of the icosahedron projection are reasonably traced, while the traditional phasing method is only able to produce an elongated shape without any detail, and a much more gradual fading into the outer areas of the support.

\begin{figure}[htbp]
\centering\includegraphics[width=12cm]{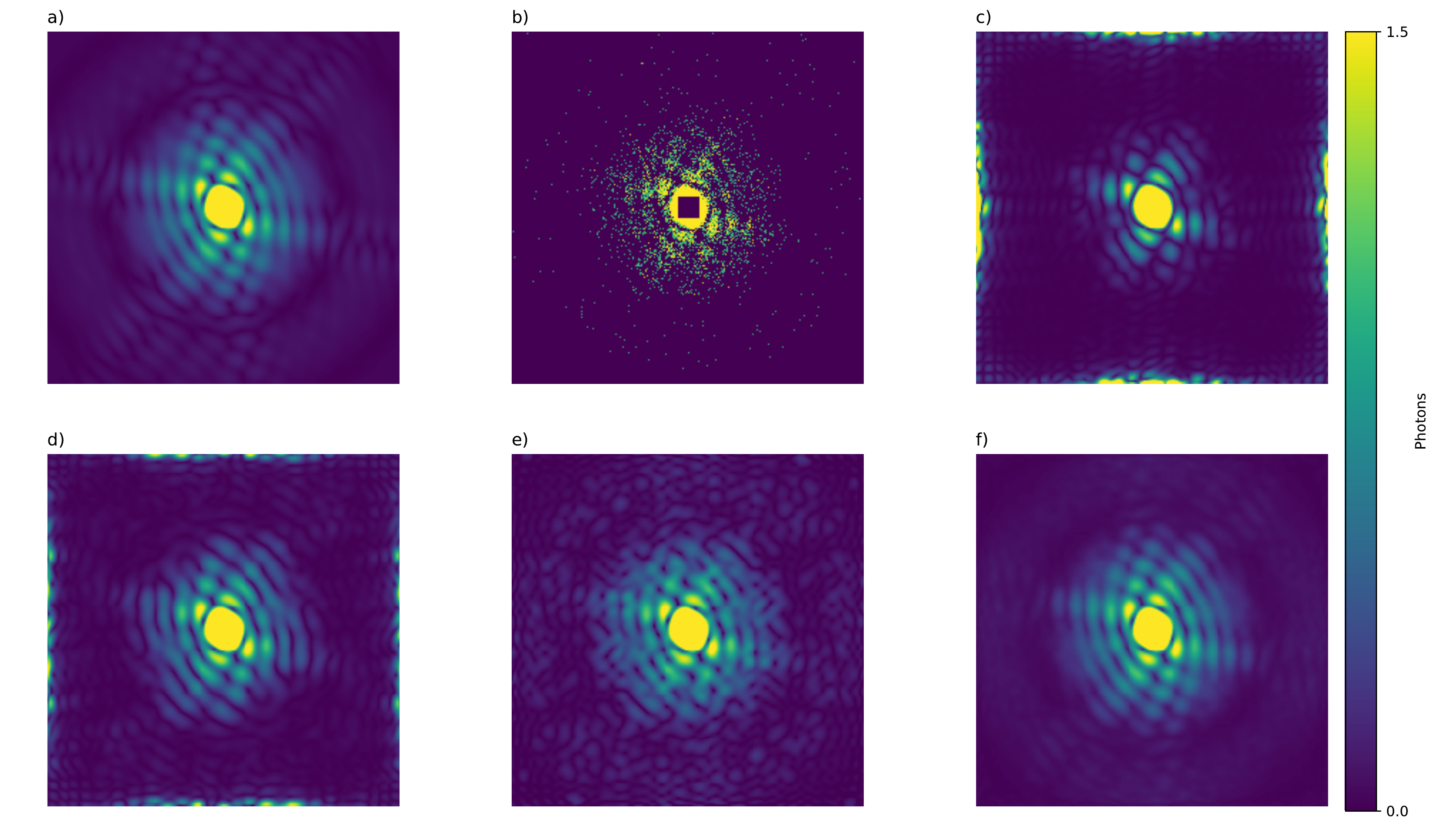}
\caption{Diffraction pattern results for the first random instance (rescaled to undo Hann windowing). a) Original simulated pattern. b) Poisson sampling with simulated 25x25 beamstop. c) Average diffraction pattern over 10 best phasings. Artifacts at the edges, reproduction of original features dropping off quickly. d) Average diffraction pattern over 10 best phasings based on COACS healing. Features reproduced to higher angles, but still problems at the edge, missing the outer ring. e) Diffraction pattern resulting directly from the COACS healing. Artifacts present in d) are absent. Outer ring corresponding to the one found in original pattern. f) Average based on the separate healing results of all 50 sampled particles. Errors cancel out, resulting in an image very similar to a).}
\label{patternexample}
\end{figure}

The diffraction pattern with various stages of processing is shown for the first random instance in Figure \ref{patternexample}. The reconstructed patterns look similar to the original in Figure \ref{cropped}. The phased COACS patterns and especially the non-phased COACS patterns are clearly superior to the patterns achieved using traditional phasing. There is no perceptible ringing effects from the highly regular rectangular support used in the COACS patterns. 

The crystallographic R factor, or equivalently the L1 norm of the difference between the recovered \emph{wave amplitudes} and the true amplitudes, normalized by the norm of the true amplitudes \cite{Rfactor}, is presented in Table \ref{rfactor}. The R factor with no healing is more than 100\% higher than what is obtained when phasing with a healing pre-processing step. However, even this result has an error significantly larger than what is obtained from the healing step itself. Thus, the phasing process is still a significant contributor to errors. Since these calculations go all the way to the edge of the detector, where the sampled pattern was exceedingly sparse, the values are relatively high.

\begin{table}[htbp]
\centering\caption{R factor means and standard deviations.}
\begin{tabular}{l|c|c}
\hline
Dataset & Avg R-factor & $\sigma$ \\ \hline
Average phasing without COACS healing & 0.416 & 0.017 \\
Average phasing with COACS healing & 0.158 & 0.019 \\
COACS healed & 0.098 & 0.005 \\
\hline
\end{tabular}
\label{rfactor}
\end{table}

\begin{figure}[htbp]
\centering\includegraphics[width=12cm]{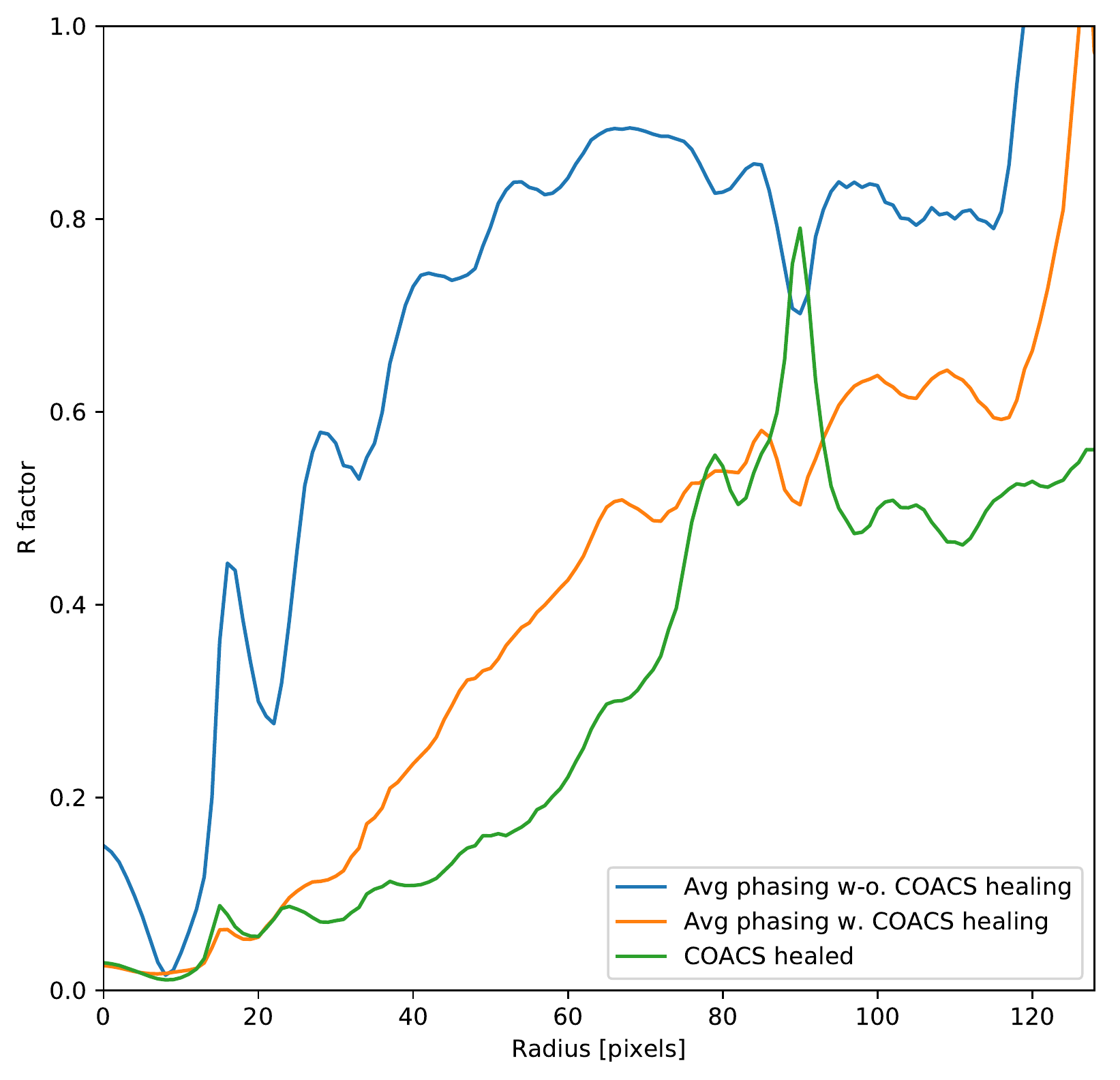}
\caption{R factors (normalized relative L1 error) for various radius shells in pixels. Curves are averages over the individually computed results for all 50 simulated particles. Comparison between results based on average phasing of 10 best reconstructions of original pattern, average phasing of 10 best reconstructions of COACS-healed pattern, and using the structural factors from the COACS pattern directly. COACS-healing reduces phasing errors, but the phasing step is still a significant contributor to errors. Peak at around 90 pixels is due to the R factor being a relative error measure. This is the location of a minimum due to the shape of the small spherical feature. Hence, absolute errors of the same magnitude are amplified.}
\label{rplot}
\end{figure}

In Figure \ref{rplot}, the R factors calculated at various radii (in pixels) are also shown. The traditional phasing method is not able to correctly recover the intensity in the central missing data region (being a square of side 25, this is seen most up to a radius of 12.5). In the overall minimum for the true signal due to the shape of the high-density spherical feature in the pattern at 90 pixels, the direct application of COACS gives inferior results due to the more specklish nature of that pattern. Although the relative error is higher at this point, the absolute error is small nonetheless, since the true signal is close to 0.

The healing process required approximately $30$ minutes per pattern on a 12-core server with Sandy Bridge Intel Xeon cores. In the Discussion section, we outline how this can be improved.

\section{Discussion}
The time usage for the intensity healing is much longer compared to phasing, 30 minutes vs. less than 10 seconds (for a single reconstruction within the batch of 100). Nonetheless, both perform roughly the same number of iterations (on the order of $10^5$). It should be noted that the current COACS implementation is a straightforward one in Matlab using the TFOCS toolbox, while the phasing implementation \texttt{libspimage} \cite{hawk} is efficient GPU-based CUDA code. When using the TFOCS mode for counting the number of operations, it is clear that the number of function and transform evaluations scales linearly to the number of iterations. The adaptive backtracking of the step-size in TFOCS does mean that the number of evaluations can be higher, but in our testing this was still lower than a total factor of 5. A preliminary analysis of the computational workload indicates that the Fourier transforms should be the dominating part of the computational load. Thus, an efficient implementation of the TFOCS algorithm should be quite similar to the existing phasing, in terms of work per iteration. The time difference is due to the difference between a CPU-based implementation in Matlab for intensity healing, and an efficient C++ implementation on GPUs using CUDA. Another avenue would be to use the existing port of TFOCS methodology into the TensorFlow framework.

We can also conclude that our current acceleration scheme is ad hoc rather than optimal. In fact, the translation of our problem into one of general convex optimization means that one can much more easily identify and benefit from existing acceleration schemes that have shown success in other applications. We have tried reformulations using the TFOCS SCD scheme with accompanying relaxations of $\mu$ (not shown), but those did not perform better than our ad hoc approach in terms of accuracy or performance.

We believe that slight additional relaxations like the ones already outlined, and an efficient GPU-based implementation, should make intensity healing achievable at time scales similar to phasing. More importantly, we have demonstrated that a proper numerical treatment allows proper intensity healing and phasing without a carefully chosen tight support or Shrinkwrap procedure. This development holds the promise of allowing routine phasing of recorded experimental diffraction patterns, without tedious manual tuning. With equally tight support constraints, it is also foreseeable that this method will allow better handling of patterns with missing data in some parts of the pattern due to e.g. saturation, the presence of a beamstop to protect the detector from the much stronger non-diffracted beam, or other aspects of detector or experiment geometry. For analyzing real-world data, one will probably want to augment our current Poisson probability distribution with a Gaussian component for values close to 0, to better account for electronic noise in imaging detectors.

Another observation based on our evaluation is that the phasing schemes are still the weakest step in the reconstruction procedure. Even the smooth COACS patterns result in reconstruction processes that start to ``walk'' along the image. This is due to a non-ideal phase ramp being induced by the combination of the two constraints, resulting in movement in real space. This, in turn, means that the real space object will repeatedly ``bump'' into the support, with new artifacts being introduced as the real space constraint implementation tries to remove the signal. While the scope of this publication is not to improve phasing per se, new insights can be gained by clearly separating effects due to sparse data, the difference between the continuous and discrete Fourier transforms, and the phasing methodology itself. We also note that a few individual reconstructions look far sharper than the averages shown, although the error metrics do not reliably separate those. It is our opinion that a proper phasing method should be able to produce R factors similar to those we report for the COACS method alone in Table \ref{rfactor} and Figure \ref{rplot}.

The straightforward structure and generality of the convex optimization formalism as well as the TFOCS library in particular also make it feasible to add additional constraints. Such constraints might include a  total variation norm in the autocorrelation or Fourier space to regularize the problem. Especially in exceedingly sparse cases, and with challenging experiment geometries, such additional inspiration from the compressed sensing literature might prove worthwhile.

\section{Conclusion}
We have presented the COACS approach to correct the sampled diffraction pattern based on a support constraint. This approach allows for higher-resolution reconstructions, which made it possible for us to phase simulated data with a wide support and no specific tuning of the parameters with a high level of detail. We have also identified several possible future developments, most pressingly to implement a GPU-based version of our approach in order to present more competitive performance in terms of computation time.

\ifthenelse{\boolean{osa}}{
\section*{Funding}
Please identify all appropriate funding sources by name and contract number. Funding information should be listed in a separate block preceding any acknowledgments.

List only the funding agencies and any associated grants or project numbers, as shown in the example below:\\
National Science Foundation (NSF) (1253236, 0868895, 1222301); Program 973 (2014AA014402); Natural National Science Foundation (NSFC) (123456).

OSA participates in http://www.crossref.org/fundingdata Crossref's Funding Data, a service that provides a standard way to report funding sources for published scholarly research. To ensure consistency, please enter any funding agencies and contract numbers from the Funding section in Prism during submission or revisions.

\subsection{Bib\TeX}
\label{sec:bibtex}
Bib\TeX{} may be used to create a file containing the references, whose contents (i.e., contents of \texttt{.bbl} file) can then be pasted into the bibliography section of the \texttt{.tex} file. A new Bib\TeX{} style file, \texttt{osajnl.bst}, is provided.

To assist authors with journal abbreviations in references, standard abbreviations for some commonly cited journals have been included as macros within opex3.sty.  The abbreviations are shown in Table 2 below.

\begin{table}[htbp]
\centering\caption{Standard abbreviations
 for commonly cited journals.}
\begin{tabular}{lp{1.7in}|lp{1.7in}}
\hline
Macro        & Abbreviation                & Macro        & Abbreviation          \\ \hline
\verb+\ao+   & Appl.\  Opt.\               & \verb+\jpp+  & J. Phys.              \\
\verb+\aop+  & Adv. Opt. Photon.           & \verb+\nat+  & Nature                \\
\verb+\ap+   & Appl.\  Phys.\              & \verb+\oc+   & Opt.\ Commun.\        \\
\verb+\apl+  & Appl.\ Phys.\ Lett.\        & \verb+\opex+ & Opt.\ Express         \\
\verb+\apj+  & Astrophys.\ J.\             & \verb+\ol+   & Opt.\ Lett.\          \\
\verb+\bell+ & Bell Syst.\ Tech.\ J.\      & \verb+\ome+  & Opt.\ Mater.\ Express \\
\verb+\boe+  & Biomed.\ Opt.\ Express      & \verb+\opn+  & Opt.\ Photon.\ News   \\
\verb+\jqe+ & IEEE J.\ Quantum Electron.\  & \verb+\pl+   & Phys.\ Lett.\         \\
\verb+\assp+ & IEEE Trans.\ Acoust.\ Speech Signal Process.\  & \verb+\pr+ & Photon.\ Res.\ \\
\verb+\aprop+ & IEEE Trans.\  Antennas Propag.\    & \verb+\pra+ & Phys.\ Rev.\ A   \\
\verb+\mtt+ & IEEE Trans.\ Microwave Theory Tech.\ & \verb+\prb+ & Phys.\ Rev.\ B   \\
\verb+\iovs+ & Invest.\ Ophthalmol.\ Vis.\ Sci.\    & \verb+\prc+ & Phys.\ Rev.\ C   \\
\verb+\jcp+ & J.\ Chem.\ Phys.\            & \verb+\prd+ & Phys.\ Rev.\ D   \\
\verb+\jmo+ & J.\ Mod.\ Opt.\              & \verb+\pre+ & Phys.\ Rev.\ E   \\
\verb+\jocn+ & J.\ Opt.\ Commun.\ Netw.\   & \verb+\prl+ & Phys.\ Rev.\ Lett.\    \\
\verb+\jon+ & J.\ Opt.\ Netw.\             & \verb+\rmp+ & Rev.\ Mod.\ Phys.\    \\
\verb+\josa+ & J.\ Opt.\ Soc.\ Am.\        & \verb+\pspie+ & Proc.\ Soc.\ Photo-Opt.\ Instrum.\ Eng.\   \\
\verb+\josaa+ & J.\ Opt.\ Soc.\ Am.\ A     & \verb+\sjqe+ & Sov.\ J.\ Quantum Electron.\   \\
\verb+\josab+ & J.\ Opt.\ Soc.\ Am.\ B     & \verb+\vr+ & Vision Res.\   \\ \hline
\end{tabular}
\end{table}
}{
\bibliography{coacs}{}

\begin{thebibliography}{10}
\newcommand{\enquote}[1]{``#1''}

\bibitem{crystalphasing}
P.~D. Adams, P.~V. Afonine, R.~W. Grosse-Kunstleve, R.~J. Read, J.~S.
  Richardson, D.~C. Richardson, and T.~C. Terwilliger, \enquote{Recent
  developments in phasing and structure refinement for macromolecular
  crystallography,} {Current Opinion in Structural Biology} \textbf{19},
  566--572 (2009).

\bibitem{hio}
J.~R. Fienup, \enquote{Phase retrieval algorithms: a comparison,} {Applied
  Optics} \textbf{21}, 2758--2769 (1982).

\bibitem{raar}
D.~R. Luke, \enquote{Relaxed averaged alternating reflections for diffraction
  imaging,} Inverse problems \textbf{21}, 37 (2004).

\bibitem{marchesini}
S.~Marchesini, \enquote{Invited article: A unified evaluation of iterative
  projection algorithms for phase retrieval,} {Review of Scientific
  Instruments} \textbf{78}, 011301 (2007).

\bibitem{lcls}
S.~Boutet and G.~J. Williams, \enquote{{The coherent X-ray imaging (CXI)
  instrument at the Linac Coherent Light Source (LCLS)},} {New Journal of
  Physics} \textbf{12}, 035024 (2010).

\bibitem{nanorice}
A.~V. Martin, J.~Andreasson, A.~Aquila, S.~Bajt, T.~R. Barends, M.~Barthelmess,
  A.~Barty, W.~H. Benner, C.~Bostedt, J.~D. Bozek \emph{et~al.},
  \enquote{Single particle imaging with soft x-rays at the linac coherent light
  source,} in \enquote{Advances in X-ray Free-Electron Lasers: Radiation
  Schemes, X-ray Optics, and Instrumentation,} , vol. 8078 (International
  Society for Optics and Photonics, 2011), vol. 8078, p. 807809.

\bibitem{flashman}
H.~N. Chapman, A.~Barty, M.~J. Bogan, S.~Boutet, M.~Frank, S.~P. Hau-Riege,
  S.~Marchesini, B.~W. Woods, S.~Bajt, W.~H. Benner \emph{et~al.},
  \enquote{{Femtosecond diffractive imaging with a soft-X-ray free-electron
  laser},} {Nature Physics} \textbf{2}, 839 (2006).

\bibitem{mimi}
M.~M. Seibert, T.~Ekeberg, F.~R. Maia, M.~Svenda, J.~Andreasson,
  O.~J{\"o}nsson, D.~Odi{\'c}, B.~Iwan, A.~Rocker, D.~Westphal \emph{et~al.},
  \enquote{{Single mimivirus particles intercepted and imaged with an X-ray
  laser},} Nature \textbf{470}, 78 (2011).

\bibitem{carboxy}
M.~F. Hantke, D.~Hasse, F.~R. Maia, T.~Ekeberg, K.~John, M.~Svenda, N.~D. Loh,
  A.~V. Martin, N.~Timneanu, D.~S. Larsson \emph{et~al.},
  \enquote{High-throughput imaging of heterogeneous cell organelles with an
  x-ray laser,} {Nature Photonics} \textbf{8}, 943 (2014).

\bibitem{phaselift1}
E.~J. Candes, T.~Strohmer, and V.~Voroninski, \enquote{Phaselift: Exact and
  stable signal recovery from magnitude measurements via convex programming,}
  {Communications on Pure and Applied Mathematics} \textbf{66}, 1241--1274
  (2013).

\bibitem{phaseliftsiam2015}
E.~J. Candes, Y.~C. Eldar, T.~Strohmer, and V.~Voroninski, \enquote{Phase
  retrieval via matrix completion,} {SIAM Review} \textbf{57}, 225--251 (2015).

\bibitem{phasecut}
I.~Waldspurger, A.~d’Aspremont, and S.~Mallat, \enquote{Phase recovery,
  maxcut and complex semidefinite programming,} {Mathematical Programming}
  \textbf{149}, 47--81 (2015).

\bibitem{tfocs}
S.~R. Becker, E.~J. Cand{\`e}s, and M.~C. Grant, \enquote{Templates for convex
  cone problems with applications to sparse signal recovery,} Mathematical
  Programming Computation \textbf{3}, 165 (2011).

\bibitem{3dmimi}
T.~Ekeberg, M.~Svenda, C.~Abergel, F.~R. Maia, V.~Seltzer, J.-M. Claverie,
  M.~Hantke, O.~J{\"o}nsson, C.~Nettelblad, G.~Van Der~Schot \emph{et~al.},
  \enquote{Three-dimensional reconstruction of the giant mimivirus particle
  with an x-ray free-electron laser,} {Physical Review Letters} \textbf{114},
  098102 (2015).

\bibitem{hawk}
F.~R. Maia, T.~Ekeberg, D.~Van Der~Spoel, and J.~Hajdu, \enquote{Hawk: the
  image reconstruction package for coherent x-ray diffractive imaging,}
  {Journal of Applied Crystallography} \textbf{43}, 1535--1539 (2010).

\bibitem{thibaulterror}
P.~Thibault, V.~Elser, C.~Jacobsen, D.~Shapiro, and D.~Sayre,
  \enquote{Reconstruction of a yeast cell from x-ray diffraction data,} {Acta
  Crystallographica A} \textbf{62}, 248--261 (2006).

\bibitem{slacpaper}
L.~Shi, G.~Wetzstein, and T.~J. Lane, \enquote{A flexible phase retrieval
  framework for flux-limited coherent x-ray imaging,} arXiv preprint
  arXiv:1606.01195  (2016).

\bibitem{interiorreview}
Y.~Nesterov, \emph{Introductory lectures on convex optimization: A basic
  course}, vol.~87 (Springer Science \& Business Media, 2013).

\bibitem{solverfree}
J.~L{\"o}fberg, \enquote{Yalmip: A toolbox for modeling and optimization in
  matlab,} in \enquote{Computer Aided Control Systems Design, 2004 IEEE
  International Symposium on,}  (IEEE, 2004), pp. 284--289.

\bibitem{solvercomm}
M.~ApS, \emph{The MOSEK optimization toolbox for MATLAB manual. Version 8.1.}
  (2017).

\bibitem{cvx}
M.~Grant, S.~Boyd, and Y.~Ye, \enquote{Cvx: Matlab software for disciplined
  convex programming,}  (2008).

\bibitem{sphere}
\emph{Absorption and Scattering by a Sphere} (Wiley-Blackwell, 2007), chap.~4,
  pp. 82--129.

\bibitem{tfocsug}
S.~Becker, E.~Candes, and M.~Grant, \emph{Templates for first-order conic
  solvers user guide 1.3} (2013).

\bibitem{condor}
M.~F. Hantke, T.~Ekeberg, and F.~R. Maia, \enquote{Condor: a simulation tool
  for flash x-ray imaging,} {Journal of Applied Crystallography} \textbf{49},
  1356--1362 (2016).

\bibitem{shrinkwrap}
S.~Marchesini, H.~He, H.~N. Chapman, S.~P. Hau-Riege, A.~Noy, M.~R. Howells,
  U.~Weierstall, and J.~C. Spence, \enquote{X-ray image reconstruction from a
  diffraction pattern alone,}  .

\bibitem{Rfactor}
W.~C. Hamilton, \enquote{Significance tests on the crystallographic r factor,}
  {Acta Crystallographica} \textbf{18}, 502--510 (1965).

\end{thebibliography}
}

\end{document}